\pdfoutput=1

\documentclass[11pt]{article}

\usepackage[]{acl}
\usepackage{tablefootnote}
\usepackage{threeparttable}
\usepackage{graphicx} 
\usepackage{times}
\usepackage{latexsym}
\usepackage{amsfonts,amssymb}
\usepackage{float}
\usepackage{array}
\usepackage{booktabs}
\usepackage{epstopdf}
\usepackage{color}
\usepackage{hyperref}
\usepackage{stfloats}
\usepackage[T1]{fontenc}

\usepackage[utf8]{inputenc}

\usepackage{microtype}

\title{The NPU-MSXF Speech-to-Speech Translation System for IWSLT 2023 Speech-to-Speech Translation Task}

\vspace{-3pt}
\author{Kun Song$^1$, Yi Lei$^1$, Peikun Chen$^1$, Yiqing Cao$^2$, Kun Wei$^1$, Yongmao Zhang$^1$,\\ {\bf Lei Xie$^{1}$$^*$, Ning Jiang$^3$, Guoqing Zhao$^3$}\\
  $^1$Audio, Speech and Language Processing Group (ASLP@NPU), \\School of Computer Science,
  Northwestern Polytechnical University, China \\
  $^2$Department of Computer Science and Technology, Nanjing University, China\\
  $^3$MaShang Consumer Finance Co., Ltd, China}

\begin{document}
\maketitle
\footnoteONLYtext{$^*$Lei Xie is the corresponding author.}
\vspace{-10pt}
\begin{abstract}
\vspace{-5pt}
This paper describes the NPU-MSXF system for the IWSLT 2023 speech-to-speech translation (S2ST) task which aims to translate from English speech of multi-source to Chinese speech. The system is built in a cascaded manner consisting of automatic speech recognition (ASR), machine translation (MT), and text-to-speech (TTS). We make tremendous efforts to handle the challenging multi-source input. Specifically, to improve the robustness to multi-source speech input, we adopt various data augmentation strategies and a ROVER-based score fusion on multiple ASR model outputs. To better handle the noisy ASR transcripts, we introduce a three-stage fine-tuning strategy to improve translation accuracy. Finally, we build a TTS model with high naturalness and sound quality, which leverages a two-stage framework, using network bottleneck features as a robust intermediate representation for speaker timbre and linguistic content disentanglement. Based on the two-stage framework, pre-trained speaker embedding is leveraged as a condition to transfer the speaker timbre in the source English speech to the translated Chinese speech. Experimental results show that our system has high translation accuracy, speech naturalness, sound quality, and speaker similarity. Moreover, it shows good robustness to multi-source data.~\footnote{Our submitted system \textbf{ranks 1st} in the S2ST task.}
\end{abstract}

\vspace{-15pt}
\section{Introduction}
\vspace{-5pt}
In this paper, we describe NPU-MSXF team's cascaded speech-to-speech translation (S2ST) system submitted to the speech-to-speech (S2S) track\footnote{\url{https://iwslt.org/2023/s2s}} of the IWSLT 2023 evaluation campaign.  The S2S track aims to build an offline system that realizes speech-to-speech translation from English to Chinese. Particularly, the track allows the use of large-scale data, including the data provided in this track as well as all training data from the offline track\footnote{\url{https://iwslt.org/2023/offline}} on speech-to-text translation task. Challengingly, the test set contains multi-source speech data, covering a variety of acoustic conditions and speaking styles, designed to examine the robustness of the S2ST system. Moreover, speaker identities conveyed in the diverse multi-source speech test data are unseen during training, which is called \textit{zero-shot S2ST} and better meets the demands of real-world applications.

Current mainstream S2ST models usually include \textit{cascaded} and \textit{end-to-end} systems. Cascaded S2ST systems, widely used in the speech-to-speech translation task~\cite{DBLP:journals/taslp/NakamuraMNKKJZYSY06}, usually contain three modules, i.e. automatic speech recognition (ASR), machine translation (MT), and text-to-speech (TTS). Meanwhile, end-to-end (E2E) S2ST systems~\cite{DBLP:conf/interspeech/JiaWBMJCW19, DBLP:conf/acl/LeeCWGPMPAHTPH22} have recently come to the stage by integrating the above modules into a unified model for directly synthesizing target language speech translated from the source language.  E2E S2ST systems can effectively simplify the overall pipeline and alleviate possible error propagation. Cascaded S2ST systems may also alleviate the error propagation problem by leveraging the ASR outputs for MT model fine-tuning. Meanwhile, thanks to the individual training process of sub-modules, cascaded systems can make better use of large-scale text and speech data, which can significantly promote the performance of each module. 


In this paper, we build a cascaded S2ST system aiming at English-to-Chinese speech translation with preserving the speaker timbre of the source English speech. The proposed system consists of Conformer-based~\cite{DBLP:conf/interspeech/GulatiQCPZYHWZW20} ASR models, a pretrain-finetune schema-based MT model~\cite{radford2018improving}, and a VITS-based TTS model~\cite{DBLP:conf/icml/KimKS21}. For ASR, model fusion and data augmentation strategies are adopted to improve the recognition accuracy and generalization ability of ASR with multi-source input. For MT, we use a three-stage fine-tuning process to adapt the translation model to better facilitate the output of ASR. Meanwhile, back translation and multi-fold verification strategies are adopted. Our TTS module is composed of a text-to-BN stage and a BN-to-speech stage, where speaker-independent neural bottleneck (BN) features are utilized as an intermediate representation bridging the two stages. Specifically, the BN-to-speech module, conditioned on speaker embedding extracted from the source speech, is to synthesize target language speech with preserving the speaker timbre. Combined with a pre-trained speaker encoder to provide speaker embeddings, the TTS model can be generalized to unseen speakers, who are not involved in the training process. Experimental results demonstrate the proposed S2ST system achieves good speech intelligibility, naturalness, sound quality, and speaker similarity. 
\vspace{-6pt}
\section{Automatic Speech Recognition}
\vspace{-5pt}
Our ASR module employs multiple models for score fusion in the inference. Moreover, data augmentation is adopted during training to handle noisy multi-source speech.

\vspace{-6pt}
\subsection{Model Structure}
\vspace{-5pt}

Our system employs both Conformer~\cite{DBLP:conf/interspeech/GulatiQCPZYHWZW20} and E-Branchformer models~\cite{kim2023branchformer} in our ASR module to address the diversity of the test set. Conformer sequentially combines convolution, self-attention, and feed-forward layers. The self-attention module serves to capture global contextual information from the input speech, while the convolution layer focuses on extracting local correlations. This model has demonstrated remarkable performance in ASR tasks with the ability to capture local and global information from input speech signals. E-Branchformer uses dedicated branches of convolution and self-attention based on the Conformer and applies efficient merging methods, in addition to stacking point-wise modules. E-Branchformer achieves state-of-the-art results in ASR.

\vspace{-6pt}
\subsection{Data Augmentation}
\vspace{-3pt}
Considering the diversity of the testing data, we leverage a variety of data augmentation strategies to expand the training data of our ASR system, including the following aspects.

\begin{itemize}
\vspace{-8pt}
\item \textbf{Speed Perturbation}: We notice that the testing set contains spontaneous speech such as conversations with various speaking speeds. So speed perturbation is adopted to improve the generalization ability of the proposed model. Speed perturbation is the process of changing the speed of an audio signal while preserving other information (including pitch) in the audio. We perturb the audio speech with a speed factor of 0.9, 1.0, and 1.1 to all the training data. Here speed factor refers to the ratio compared to the original speed of speech.
\vspace{-6pt}
\item \textbf{Pitch Shifting}: Pitch shifting can effectively vary the speaker identities to increase data diversity. Specifically, we use SoX\footnote{\url{https://sox.sourceforge.net/}} audio manipulation tool to perturb the pitch in the range [-40, 40].
\vspace{-6pt}
\item \textbf{Noise Augmentation}: There are many cases with heavy background noise in the test set, including interfering speakers and music. However, the data set provided by the organizer is much cleaner than the test set, which makes it necessary to augment the training data by adding noises to improve the recognition performance. Since there is no noise set available, we create a noise set from the data provided. A statistical VAD~\cite{DBLP:journals/spl/SohnKS99} is used to cut the non-vocal and vocal segments from the data and the non-vocal segments with energy beyond a threshold comprise our noise set. We add the noise segments to the speech utterances with a signal-to-noise ratio ranging from 0 to 15 dB.
\vspace{-6pt}
\item \textbf{Audio Codec}: Considering the test data come from multiple sources, we further adopt audio codec augmentation to the training data. Specifically, we use the FFmpeg\footnote{\url{https://ffmpeg.org/}} tool to convert the original audio to Opus format at [48, 96, 256] Kbps. 
\vspace{-6pt}
\item \textbf{Spectrum Augmentation}: 
To prevent the ASR model from over-fitting, we apply the SpecAugment method~\cite{DBLP:conf/interspeech/ParkCZCZCL19} to the input features during every mini-batch training. SpecAugment includes time warping, frequency channel masking, and time step masking, and we utilize all of these techniques during training.
\end{itemize}
\vspace{-10pt}
\subsection{Model Fusion}
\vspace{-5pt}
Since a single ASR model may overfit to a specific optimization direction during training, it cannot guarantee good recognition accuracy for the speech of various data distributions. To let the ASR model generalize better to the multi-source input, we adopt a model fusion strategy. 
Specifically, we train the Conformer and E-branchformer models introduced in Section 2.1 using the combination of the original and the augmented data. Each testing utterance is then transcribed by these different models, resulting in multiple outputs. Finally, ROVER~\cite{fiscus1997post} is adopted to align and vote with equal weights on the multiple outputs, resulting in the final ASR output.
\vspace{-8pt}
\subsection{ASR Output Post-processing}
\vspace{-5pt}
Given that the spontaneous speech in the test set contains frequent filler words such as "Uh" and "you know", it is necessary to address their impact on subsequent MT accuracy and TTS systems that rely on the ASR output. To mitigate this issue, we use a simple rule-based post-processing step to detect and eliminate these expressions from the ASR output. By doing so, we improve the accuracy of the downstream modules.
\vspace{-8pt}
\section{Machine Translation}
\vspace{-5pt}
For the MT module, we first use a pre-trained language model as a basis for initialization and then employ various methods to further enhance translation accuracy.
\vspace{-8pt}
\subsection{Pre-trained Language Model}
\vspace{-5pt}

As pre-trained language models are considered part of the training data in the offline track and can be used in the S2ST track, we use the pre-trained mBART50 model for initializing our MT module. mBART50~\cite{DBLP:journals/tacl/LiuGGLEGLZ20} is a multilingual BART~\cite{DBLP:conf/acl/LewisLGGMLSZ20} model with 12 layers of encoder and decoder, which we believe will provide a solid basis for improving translation accuracy.

\vspace{-8pt}
\subsection{Three-stage Fine-tuning Based on Curriculum Learning}
\vspace{-5pt}

We perform fine-tuning on the pre-trained model to match the English-to-Chinese (En2Zh) translation task. There are substantial differences between the ASR outputs and the texts of MT data. First, ASR prediction results inevitably contain errors. Second, ASR outputs are normalized text without punctuation.
Therefore, directly fine-tuning the pre-trained model with the MT data will cause a mismatch problem with the ASR output during inference. On the other hand, fine-tuning the model with the ASR outputs will cause difficulty in model coverage because of the difference between the ASR outputs and the MT data. Therefore, based on Curriculum Learning~\cite{DBLP:conf/icml/BengioLCW09}, we adopt a three-stage fine-tuning strategy to mitigate such a mismatch.
\begin{itemize}
\vspace{-6pt}
\item \textbf{Fine-tuning using the MT data}: First, we use all the MT data to fine-tune the pre-trained model to improve the accuracy of the model in the En2Zh translation task.
\vspace{-6pt}
\item \textbf{Fine-tuning using the MT data in ASR transcription format}: Second, we convert the English text in the MT data into the ASR transcription format. Then, we fine-tune the MT model using the converted data, which is closer to the actual text than the ASR recognition output. This approach can enhance the stability of the fine-tuning process, minimize the impact of ASR recognition issues on the translation model, and improve the model's ability to learn punctuation, thereby enhancing its robustness.
\vspace{-6pt}
\item \textbf{Fine-tuning using the ASR outputs}: Third, we leverage \textit{GigaSpeech}~\cite{DBLP:conf/interspeech/ChenCWDZWSPTZJK21} to address the mismatch problem between the ASR outputs and the MT data. Specifically, we use the ASR module to transcribe the \textit{GigaSpeech} training set and replace the corresponding transcriptions in \textit{GigaST}~\cite{DBLP:journals/corr/abs-2204-03939} with the ASR transcriptions for translation model fine-tuning. This enables the MT model to adapt to ASR errors.
\vspace{-3pt}
\end{itemize}

\vspace{-8pt}
\subsection{Back Translation}
\vspace{-5pt}
Following~\cite{DBLP:conf/wmt/AkhbardehABBCCC21}, we adopt the back translation method to enhance the data and improve the robustness and generalization of the model. First, we train a Zh2En MT model to translate Chinese to English, using the same method employed for the En2Zh MT module. Next, we generate the corresponding English translations for the Chinese text of the translation data. Finally, we combine the back translation parallel corpus pairs with the real parallel pairs and train the MT model.

\vspace{-8pt}
\subsection{Cross-validation} 
\vspace{-5pt}
We use 5-fold cross-validation~\cite{DBLP:journals/jmlr/OjalaG10} to improve the robustness of translation and reduce over-fitting. Firstly, we randomly divide the data into five equal parts and train five models on different datasets by using one of them as the validation set each time and combining the remaining four as the training set. After that, we integrate the predicted probability distributions from these five models to obtain the final predicted probability distribution for the next word during token generation for predicting the translation results.

\vspace{-8pt}
\section{Text-to-speech}
\vspace{-5pt}
\subsection{Overview}
\vspace{-5pt}

\begin{figure*}[ht]

	\centering
	\includegraphics[width=16.8cm,height=6.3cm]{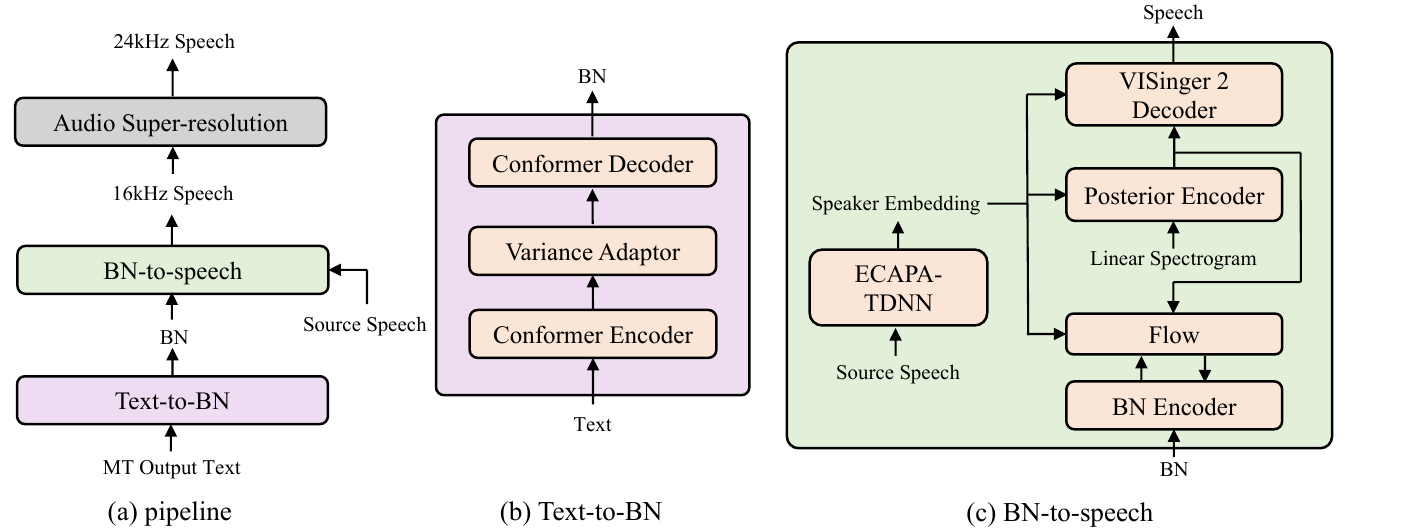}
	\vspace{-15pt}
        \caption{Architecture of our text-to-speech module.}
       
	\label{model_tts}
 \vspace{-5pt}
\end{figure*}

Figure~\ref{model_tts} (a) shows the pipeline of the text-to-speech module in the proposed S2ST system. The TTS module is built on a BN-based two-stage architecture, which consists of a text-to-BN and a BN-to-speech procedure. The text-to-BN stage tends to generate BN features from the Chinese text translated by the MT module. The BN-to-speech stage produces 16KHz Chinese speech from the BN feature, conditioning on the speaker embedding of source speech. Given the translated Chinese speech which preserves the speaker timbre in the source English speech, an audio super-resolution model is further leveraged to convert the synthesized speech from 16KHz to 24KHz for higher speech fidelity.


Building on the two-stage framework AdaVITS~\cite{DBLP:conf/iscslp/SongXWCZXYZS22}, we employ bottleneck (BN) features as the intermediate representations in the two-stage TTS module. BN features, extracted from a multi-condition trained noise-robust ASR system, mainly represent the speaker-independent linguistic content. So BN can effectively \textit{disentangle} the speaker timbre and the linguistic content information. 
In the text-to-BN stage, high-quality TTS data is adopted in the training phase to model the speaker-independent BN features with prosody information. In the BN-to-speech stage, both high-quality TTS data and low-quality ASR data should be involved during training to sufficiently model the speech of various speaker identities. Extracted from speech, BN features contain the duration and prosody information, which eliminates the need for text transcripts and prosody modeling. Instead, the BN-to-speech stage focuses on time-invariant information modeling, such as speaker timbre.


As the goal of this work is to conduct zero-shot English-to-Chinese speech translation, we concentrate on the method to transfer the unseen speaker timbre of the source English speech to the synthesized Chinese speech through voice cloning~\cite{DBLP:conf/iclr/ChenASBRZWCTLGO19}. To capture new speaker timbre during inference, the TTS module requires to model abundant various speakers during training, which relies on large-scale high-quality TTS data.
Unfortunately, we are limited in the high-quality TTS data we can use in this task and must rely on additional data such as ASR to model the speaker timbre. However, this data is not suitable for TTS model training because the labels are inconsistent with TTS, and the prosody of the speakers is not as good as high-quality TTS data.

Furthermore, we incorporate ASR data into the BN-to-speech training procedure by re-sampling all the training speech to 16kHz, which can not reach high-quality audio. Therefore, we utilize audio super-resolution techniques to upsample the synthesized 16KHz audio and convert it into higher sampling rate audio.

\vspace{-8pt}
\subsection{Text-to-BN}
\vspace{-5pt}
Our text-to-BN stage network in TTS is based on DelightfulTTS~\cite{DBLP:journals/corr/abs-2110-12612}, which employs a Conformer-based encoder, decoder, and a variance adapter for modeling duration and prosody. The model extends phoneme-level linguistic features to frame-level to guarantee the clarity and naturalness of speech in our system.

\vspace{-8pt}
\subsection{BN-to-speech}
\vspace{-5pt}
We build the BN-to-speech model based on VITS~\cite{DBLP:conf/icml/KimKS21}, which is a mainstream end-to-end TTS model. VITS generates speech waveforms directly from the input textual information, rather than a conventional pipeline of using the combination of an acoustic model and a neural vocoder. 



The network of the BN-to-speech stage consists of a BN encoder, posterior encoder, decoder, flow, and speaker encoder. The monotonic alignment search (MAS) from the original VITS is removed since BN features contain the duration information. For achieving zero-shot voice cloning, an ECAPA-TDNN~\cite{DBLP:conf/interspeech/DesplanquesTD20} speaker encoder is pre-trained to provide the speaker embedding as the condition of the synthesized speech. To avoid periodic signal prediction errors in the original HiFiGAN-based~\cite{DBLP:conf/nips/KongKB20} decoder in VITS, which induces sound quality degradation, we follow VISinger2~\cite{DBLP:journals/corr/abs-2211-02903} to adopt a decoder with the sine excitation signals. Since the VISinger2 decoder requires pitch information as input, we utilize a pitch predictor with a multi-layer Conv1D that predicts the speaker-dependent pitch from BN and speaker embedding. With the desired speaker embedding and corresponding BN features, the BN-to-speech module produces Chinese speech in the target timbre.


\vspace{-10pt}
 \subsection{Audio Super-resolution}
 \vspace{-5pt}

Following~\cite{DBLP:journals/corr/abs-2110-12612}, we use an upsampling network based vocoder to achieve audio super-resolution (16kHz$\rightarrow$24kHz). During training, the 16KHz mel-spectrogram is used as the condition to predict the 24KHz audio in the audio super-resolution model. Specifically, we adopt the \textit{AISHELL-3}~\cite{DBLP:conf/interspeech/ShiBXZL21} dataset, composing the paired 16KHz and 24KHz speech data for model training. During inference, the high-quality 24kHz speech is produced for the mel-spectrogram of the 16KHz speech generated by the BN-to-speech model. Here DSPGAN~\cite{DBLP:journals/corr/abs-2211-01087} is adopted as our audio super-resolution model, which is a universal vocoder that ensures robustness and good sound quality without periodic signal errors.

\vspace{-8pt}
\section{Data Preparation}
\vspace{-5pt}
\subsection{Datasets}
\vspace{-6pt}
Following the constraint of data usage, the training dataset for the S2ST system is illustrated in Table~\ref{dataset}.

\begin{table*}[tp]
\centering
    \vspace{-5pt}
	\caption{Datasets used in our proposed system.}
\vspace{-5pt}
\begin{tabular}{lcc}
    \toprule
 Datasets  & Utterances  & Hours \\ \midrule
\midrule
\textbf{\textit{English Labeled Speech Data}} & & \\
\midrule
GigaSpeech~\cite{DBLP:conf/interspeech/ChenCWDZWSPTZJK21} & 8,315K & 10,000\\
LibriSpeech~\cite{DBLP:conf/icassp/PanayotovCPK15} &  281K & 961 \\
TED-LIUM v2~\cite{DBLP:conf/lrec/RousseauDE12}\&v3~\cite{DBLP:conf/specom/HernandezNGTE18} & 
  361K& 661 \\
  CommonVoice~\cite{DBLP:conf/lrec/ArdilaBDKMHMSTW20} &  1,225K &  1,668 \\
\midrule
\midrule
\textbf{\textit{Text-parallel Data}} & & \\
\midrule 
News Commentary~\cite{DBLP:conf/interspeech/ChenCWDZWSPTZJK21} & 322K & - \\
OpenSubtitles2018~\cite{DBLP:conf/lrec/LisonTK18} & 10M & - \\
\midrule 
\midrule
\textbf{\textit{ST Data}} & & \\
\midrule
GigaST~\cite{DBLP:journals/corr/abs-2204-03939} & 7,651K & 9,781 \\
\midrule
\midrule
\textbf{\textit{S2S Data}} & & \\
\midrule
GigaS2S\footnotemark{} & 7,626K & - \\
\midrule
\midrule
\textbf{\textit{Chinese TTS Data}} & & \\
\midrule
AISHELL-3~\cite{DBLP:conf/interspeech/ShiBXZL21} &88K & 85 \\
\bottomrule
\end{tabular}

\vspace{-10pt}
\label{dataset}
\renewcommand{\thefigure}{1}
\end{table*}
\footnotetext{\url{https://github.com/SpeechTranslation/GigaS2S}}
\vspace{-8pt}

\subsubsection{ASR Data}
\vspace{-5pt}
For the English ASR module in our proposed system, we use \textit{GigaSpeech}, \textit{LibriSpeech}, \textit{TED-LIUM v2\&v3} as training data. For the ASR system used to extract BN features in TTS, we use text-to-speech data in \textit{AISHELL-3} and Chinese speech in \textit{GigaS2S}, along with the corresponding Chinese text in \textit{GigaST}, as the training set. Since the test set's MT output text is a mix of Chinese and English, including names of people and places, the TTS module needs to support both languages. Therefore, we also add the aforementioned English data to the training set.
\vspace{-5pt}
\subsubsection{MT Data}
\vspace{-5pt}
We use the text-parallel data including \textit{News Commentary} and \textit{OpenSubtitles2018} as MT training set. Moreover, we also add the Chinese texts in \textit{GigaST} and the English texts in \textit{GigaSpeech} corresponding to the Chinese texts in \textit{GigaST} to the training set.

\vspace{-5pt}
\subsubsection{TTS Data}
\vspace{-5pt}
We use \textit{AISHELL-3} as training data in Text-to-BN and audio super-resolution. For the pre-trained speaker encoder, we adopt \textit{LibriSpeech}, which contains 1166 speakers, as the training data.
For the BN-to-speech model, in addition to using \textit{AISHELL-3} which has 218 speakers, we also use \textit{LibriSpeech} to meet the data amount and speaker number requirements of zero-shot TTS.

\vspace{-5pt}
\subsection{Data Pre-processing}
\vspace{-3pt}
\subsubsection{ASR Data}
\vspace{-5pt}

To prepare the ASR data, we pre-process all transcripts to remove audio-related tags. Next, we map the text to the corresponding byte-pair encoding (BPE) unit and count the number of BPE units in the ASR dictionary, which totals 5,000 units. For audio processing, we use a frame shift of 10ms and a frame length of 25ms and normalize all audio to 16KHz.
\vspace{-5pt}
\subsubsection{MT Data}
\vspace{-5pt}
For the MT data, we use the same tokenizer as mBART50 to perform sub-word segmentation for English and Chinese texts and to organize them into a format for neural network training. By doing so, we can maximize the benefits of initializing our translation model with mBART50 pre-trained model parameters. The mBART tokenizer mentioned above is a Unigram tokenizer. A Unigram model is a type of language model that considers each token to be independent of the tokens before it. What’s more, the tokenizer has a total of 250,054 word segmentations, supports word segmentation processing for English, Chinese, and other languages, and uses special tokens like <s>, </s>, and <unk>.
\vspace{-5pt}
\subsubsection{TTS Data}
\vspace{-5pt}
For \textit{AISHELL-3}, we downsample it to 16KHz and 24KHz respectively as the TTS modeling target and the audio super-resolution modeling target. All other data is down-sampled to 16KHz. All data in TTS adopts 12.5ms frame shift and 50ms frame length.

\textbf{Speech Enhancement.} 
Given the presence of substantial background noise in the test set, the discriminative power of speaker embeddings is significantly reduced, thereby impeding the performance of the TTS module. Furthermore, the ASR data incorporated during the training of the BN-to-speech model is also subject to background noise. Therefore, we employ a single-channel wiener filtering method ~\cite{lim1979enhancement} to remove such noise from these data. Please note that we do not perform speech enhancement on the test set in the ASR module, because there is a mismatch between the denoised audio and which is used in ASR training, and denoising will reduce the speech recognition accuracy.

\subsubsection{Evaluation Data}
For all evaluations, we use the English-Chinese (En-Zh) development data divided by the organizer from \textit{GigaSpeech}, \textit{GigaST} and \textit{GigaS2S}, including 5,715 parallel En-Zh audio segments, and their corresponding En-Zh texts. It is worth noting that the development data for evaluations has been removed from the training dataset.

\vspace{-8pt}
\section{Experiments}
\vspace{-3pt}
\subsection{Experimental Setup}

\vspace{-5pt}
All the models in our system are trained on 8 A100 GPUs and optimized with Adam~\cite{DBLP:journals/corr/KingmaB14}.

\textbf{ASR Module.} All ASR models are implemented in ESPnet\footnote{\url{https://github.com/espnet/espnet}}. Both Conformer and E-Branchformer models employ an encoder with 17 layers and a feature dimension of 512, with 8 heads in the self-attention mechanism and an intermediate hidden dimension of 2048 for the FFN. In addition, we employ a 6-layer Transformer decoder with the same feature hidden dimension as the encoder. The E-Branchformer model uses a cgMLP with an intermediate hidden dimension of 3072. The total number of parameters for the Conformer and E-Branchformer model in Section 2.1 is 147.8M and 148.9M respectively. We train the models with batch size 32 sentences per GPU for 40 epochs, and set the learning rate to 0.0015, the warm-up step to 25K.

For data augmentation, we conduct speed perturbation, pitch shifting, and audio codec on the original recordings. Spectrum augmentation and noise augmentation are used for on-the-fly model training. 

\textbf{MT Module.} All MT models are implemented in HuggingFace\footnote{\url{https://github.com/huggingface/transformers}}. Using MT data, we fine-tune the mBART-50 large model, which has 611M parameters, with a batch size of 32 sentences per GPU for 20 epochs. The learning rate is set to 3e-5 and warmed up for the first 10\% of updates and linearly decayed for the following updates. For fine-tuning using the MT data in ASR transcription format and the ASR outputs, we also fine-tune the model with batch size 32 sentences per GPU for 5 epochs and set the learning rate to 3e-5, which is warmed up for the first 5\% of updates and linearly decayed for the following updates.

\textbf{TTS Module.} 
We complete our system based on VITS official code\footnote{\url{https://github.com/jaywalnut310/vits}}. The text-to-BN follows the configuration of DelightfulTTS and has about 64M parameters. To extract the duration required for text-to-BN, we train a Kaldi\footnote{\url{https://github.com/kaldi-asr/kaldi}} model using \textit{AISHELL-3}. The ASR system used for extracting BN is the Chinese-English ASR model mentioned in Section 5.1.1. For BN-to-speech, we use a 6-layer FFT as the BN encoder and follow the other configuration in VIsinger2 with about 45M parameters in total. The pitch predictor has 4 layers of Conv1D with 256 channels. Pitch is extracted by Visinger2 decoder and DSPGAN from Harvest~\cite{DBLP:conf/interspeech/Morise17} with Stonemask. To predict pitch in DSPGAN, we use the method described in Section 4.3. Up-sampling factors in DSPGAN is set as [5, 5, 4, 3] and other configuration of DSPGAN-mm is preserved for audio super-resolution. The DSPGAN model has about 9M parameters in total. We train all the above models with a batch size of 64 sentences per GPU for 1M steps and set the learning rate to 2e-4. For the pre-trained speaker encoder, we follow the model configuration and training setup of ECAPA-TDNN (C=1024) with 14.7M parameters. 

\vspace{-5pt}
\subsection{Evaluation Models}
\vspace{-3pt}
\textbf{Baseline.} To evaluate the effectiveness of the proposed cascaded S2ST system, we adopt the original cascaded S2ST system as a baseline, including an E-Branchformer ASR model, a mBART50 MT model fine-tuned using the MT data, and an end-to-end TTS model based on VITS trained with \textit{AISHELL-3}.

\textbf{Proposed system \& Ablation study.} 
We further conduct ablation studies to evaluate each component in the proposed system. Specifically, the ablation studies are designed to verify the effectiveness of model fusion and data augmentation in ASR, three-stage fine-tuning, back translation, cross-verification in MT, two-stage training with BN, pre-trained speaker embedding, and audio super-resolution in TTS.

\subsection{Results \& Analysis}
\vspace{-1pt}
We conduct experiments on the effectiveness of each sub-module and the performance of our proposed cascaded S2ST system.

\subsubsection{ASR Module}
\vspace{-1pt}
We calculate the word error rate (WER) of each ASR module to evaluate the English speech recognition accuracy. As shown in Table~\ref{test_asr}, the WER of the proposed system has a significant drop compared with the baseline, which indicates that the proposed system greatly improves the recognition accuracy. Moreover, the results of the ablation study demonstrate the effectiveness of both model fusion and data augmentation in improving speech recognition accuracy. 
\vspace{-5pt}
\begin{table}[ht]
    
	\centering
	\caption{The WER results of each ASR module.}
        \vspace{-3pt}
	\resizebox{0.85\linewidth}{!}{
	\begin{tabular}{lcl}
		\toprule
		Model &WER (\%)  \\ 
		\midrule
		Baseline   & 13.53  \\
            Proposed system  & 10.25 \\
        \ w/o model fusion    & 11.95 \\
        \ w/o data augmentation   & 12.40 \\

        
		\bottomrule
	\end{tabular}
	}
	\label{test_asr}
	
\end{table}

\vspace{-5pt}
\subsubsection{MT Module}
\vspace{-3pt}
We evaluate our MT module in terms of the BLEU score, which measures the $n$-gram overlap between the predicted output and the reference sentence.
\begin{table}[ht]
    
	\centering
	\caption{The BLEU results of each MT module.}
        \vspace{-5pt}
	\resizebox{0.87\linewidth}{!}{
	\begin{tabular}{lcl}
		\toprule
		Model &BLEU  \\ 
		\midrule
		Baseline & 28.1 \\
            Proposed system & 33.4 \\
        \ w/o three-stage fine-tuning   & 28.7 \\
        \ w/o back translation   & 30.8 \\
        \ w/o cross-validation   & 31.0 \\

        
		\bottomrule
	\end{tabular}
	}
	\label{test_tts}
	
\end{table}

As shown in Table~\ref{test_mt}, the proposed system with three-stage fine-tuning achieves a significantly better BLEU score than the baseline, demonstrating the effectiveness of curriculum learning in our scenario. Furthermore, by incorporating back translation and cross-validation, the translation performance can be further improved.

\subsubsection{TTS Module}


We calculate the character error rate (CER) to evaluate the clarity of speech for each TTS module. The ASR system used for calculating CER is the Chinese-English ASR model mentioned in Section 5.1.1. Additionally, we conduct mean opinion score (MOS) tests with ten listeners rating each sample on a scale of 1 (worst) to 5 (best) to evaluate naturalness, sound quality, and speaker similarity.

In the ablation study without pre-trained speaker embedding, speaker ID is to control the speaker timbre of the synthesized speech. To eliminate the influence of ASR and MT results on TTS evaluation, we use the Chinese text in the evaluation data and its corresponding English source speech as the reference of speaker timbre as the test set for TTS evaluation.

As shown in Table~\ref{test_tts}, our proposed system has achieved significant improvement in naturalness, sound quality, speaker similarity, and clarity of speech compared with the baseline. Interestingly, the system without pre-trained speaker embedding has better sound quality than both the proposed system and recording. We conjecture the reason is that the pre-trained speaker embedding greatly influences the sound quality in the zero-shot TTS setup. Therefore, the quality of the synthesized 24KHz audio is superior to the 16KHz recording, which can be demonstrated by the 3.64 MOS score of the system without audio super-resolution. Meanwhile, the speaker similarity MOS score is very low due to the lack of generalization ability to unseen speakers. 
Without using the BN-based two-stage model, the system decreases performance on all indicators, which shows the effectiveness of BN as an intermediate representation in our experimental scenario.
\begin{table*}[tp]
    
	\centering
	\caption{Experimental results of TTS in terms of MOS and WER. BN means using two-stage training with BN and pre-trained spkr. embed. means using pre-trained speaker embedding. }
        \vspace{-5pt}
	\resizebox{\linewidth}{!}{
	\begin{tabular}{lccccc}
		\toprule
		Model & Clarity in CER (\%) & Naturalness (MOS)  & Sound Quality 
 (MOS) & Speaker Similarity (MOS) \\ 
		\midrule
		Baseline &7.14 &3.38$\pm$0.05 &  3.81$\pm$0.04 & 2.12$\pm$0.06 \\
            Proposed system &6.12 & 3.70$\pm$0.06 & 3.86$\pm$0.06 & 3.72$\pm$0.06 \\
        \ w/o BN  & 7.12 & 3.40$\pm$0.04 & 3.81$\pm$0.05 & 3.10$\pm$0.07 \\
        \ w/o Pre-trained spkr. embd.  & - & - & 4.05$\pm$0.05 & 2.22$\pm$0.06 \\
        \ w/o Audio super-resolution & - & -  & 3.64$\pm$0.04 & - \\

        		\midrule
		Recording   & 4.53 & 4.01$\pm$0.04  & 3.89$\pm$0.03 & 4.35$\pm$0.05 \\
		\bottomrule
	
	\end{tabular}
	}
	\label{test_mt}
	
\end{table*}
\subsubsection{System Evaluation}

Finally, we calculate the ASR-BLEU score for the baseline and the proposed system to evaluate the speech-to-speech translation performance. Specifically, we use the ASR system to transcribe the Chinese speech generated by TTS, and then compute the BLEU scores of the ASR-decoded text with respect to the reference English translations. The ASR system for transcribing Chinese speech is the same as that in Section 6.2.3. 

\begin{table}[ht]
    
	\centering
	\caption{The ASR-BLEU results of each system.}
	\resizebox{0.7\linewidth}{!}{
	\begin{tabular}{lcl}
		\toprule
		Model &ASR-BLEU  \\ 
		\midrule
		Baseline & 27.5 \\
            Proposed system & 32.2 \\

        
		\bottomrule
	\end{tabular}
	}
	\label{test_sys}
	\vspace{-3pt}
\end{table}

As shown in Table~\ref{test_sys}, our proposed system achieves a higher ASR-BLEU score than the baseline, which indicates that our proposed system has good speech-to-speech translation accuracy.

\section{Conclusion}


This paper describes the NPU-MSXF speech-to-speech translation system, which we develop for the IWSLT 2023 speech-to-speech translation task. Our system is built as a cascaded system that includes ASR, MT, and TTS modules. To ensure good performance with multi-source data, we improved each module using various techniques such as model fusion and data augmentation in the ASR, three-stage fine-tuning, back translation, and cross-validation in the MT, and two-stage training, pre-trained speaker embedding, and audio super-resolution in the TTS. Through extensive experiments, we demonstrate that our system achieves high translation accuracy, naturalness, sound quality, and speaker similarity with multi-source input.

\clearpage

\bibliography{anthology,custom}

\begin{thebibliography}{33}
\expandafter\ifx\csname natexlab\endcsname\relax\def\natexlab#1{#1}\fi

\bibitem[{Akhbardeh et~al.(2021)Akhbardeh, Arkhangorodsky, Biesialska, Bojar,
  Chatterjee, Chaudhary, Costa{-}juss{\`{a}}, Espa{\~{n}}a{-}Bonet, Fan,
  Federmann, Freitag, Graham, Grundkiewicz, Haddow, Harter, Heafield, Homan,
  Huck, Amponsah{-}Kaakyire, Kasai, Khashabi, Knight, Kocmi, Koehn, Lourie,
  Monz, Morishita, Nagata, Nagesh, Nakazawa, Negri, Pal, Tapo, Turchi, Vydrin,
  and Zampieri}]{DBLP:conf/wmt/AkhbardehABBCCC21}
Farhad Akhbardeh, Arkady Arkhangorodsky, Magdalena Biesialska, Ondrej Bojar,
  Rajen Chatterjee, Vishrav Chaudhary, Marta~R. Costa{-}juss{\`{a}}, Cristina
  Espa{\~{n}}a{-}Bonet, Angela Fan, Christian Federmann, Markus Freitag, Yvette
  Graham, Roman Grundkiewicz, Barry Haddow, Leonie Harter, Kenneth Heafield,
  Christopher Homan, Matthias Huck, Kwabena Amponsah{-}Kaakyire, Jungo Kasai,
  Daniel Khashabi, Kevin Knight, Tom Kocmi, Philipp Koehn, Nicholas Lourie,
  Christof Monz, Makoto Morishita, Masaaki Nagata, Ajay Nagesh, Toshiaki
  Nakazawa, Matteo Negri, Santanu Pal, Allahsera~Auguste Tapo, Marco Turchi,
  Valentin Vydrin, and Marcos Zampieri. 2021.
\newblock \href {https://aclanthology.org/2021.wmt-1.1} {Findings of the 2021
  conference on machine translation {(WMT21)}}.
\newblock In \emph{Proceedings of the Sixth Conference on Machine Translation,
  WMT@EMNLP 2021, Online Event, November 10-11, 2021}, pages 1--88. Association
  for Computational Linguistics.

\bibitem[{Ardila et~al.(2020)Ardila, Branson, Davis, Kohler, Meyer, Henretty,
  Morais, Saunders, Tyers, and Weber}]{DBLP:conf/lrec/ArdilaBDKMHMSTW20}
Rosana Ardila, Megan Branson, Kelly Davis, Michael Kohler, Josh Meyer, Michael
  Henretty, Reuben Morais, Lindsay Saunders, Francis~M. Tyers, and Gregor
  Weber. 2020.
\newblock Common voice: {A} massively-multilingual speech corpus.
\newblock In \emph{Proceedings of The 12th Language Resources and Evaluation
  Conference, {LREC} 2020, Marseille, France, May 11-16, 2020}, pages
  4218--4222. European Language Resources Association.

\bibitem[{Bengio et~al.(2009)Bengio, Louradour, Collobert, and
  Weston}]{DBLP:conf/icml/BengioLCW09}
Yoshua Bengio, J{\'{e}}r{\^{o}}me Louradour, Ronan Collobert, and Jason Weston.
  2009.
\newblock Curriculum learning.
\newblock In \emph{Proceedings of the 26th Annual International Conference on
  Machine Learning, {ICML} 2009, Montreal, Quebec, Canada, June 14-18, 2009},
  volume 382 of \emph{{ACM} International Conference Proceeding Series}, pages
  41--48. {ACM}.

\bibitem[{Chen et~al.(2021)Chen, Chai, Wang, Du, Zhang, Weng, Su, Povey, Trmal,
  Zhang, Jin, Khudanpur, Watanabe, Zhao, Zou, Li, Yao, Wang, You, and
  Yan}]{DBLP:conf/interspeech/ChenCWDZWSPTZJK21}
Guoguo Chen, Shuzhou Chai, Guan{-}Bo Wang, Jiayu Du, Wei{-}Qiang Zhang, Chao
  Weng, Dan Su, Daniel Povey, Jan Trmal, Junbo Zhang, Mingjie Jin, Sanjeev
  Khudanpur, Shinji Watanabe, Shuaijiang Zhao, Wei Zou, Xiangang Li, Xuchen
  Yao, Yongqing Wang, Zhao You, and Zhiyong Yan. 2021.
\newblock Gigaspeech: An evolving, multi-domain {ASR} corpus with 10, 000 hours
  of transcribed audio.
\newblock In \emph{Interspeech 2021, 22nd Annual Conference of the
  International Speech Communication Association, Brno, Czechia, 30 August - 3
  September 2021}, pages 3670--3674. {ISCA}.

\bibitem[{Chen et~al.(2019)Chen, Assael, Shillingford, Budden, Reed, Zen, Wang,
  Cobo, Trask, Laurie, G{\"{u}}l{\c{c}}ehre, van~den Oord, Vinyals, and
  de~Freitas}]{DBLP:conf/iclr/ChenASBRZWCTLGO19}
Yutian Chen, Yannis~M. Assael, Brendan Shillingford, David Budden, Scott~E.
  Reed, Heiga Zen, Quan Wang, Luis~C. Cobo, Andrew Trask, Ben Laurie,
  {\c{C}}aglar G{\"{u}}l{\c{c}}ehre, A{\"{a}}ron van~den Oord, Oriol Vinyals,
  and Nando de~Freitas. 2019.
\newblock Sample efficient adaptive text-to-speech.
\newblock In \emph{7th International Conference on Learning Representations,
  {ICLR} 2019, New Orleans, LA, USA, May 6-9, 2019}. OpenReview.net.

\bibitem[{Desplanques et~al.(2020)Desplanques, Thienpondt, and
  Demuynck}]{DBLP:conf/interspeech/DesplanquesTD20}
Brecht Desplanques, Jenthe Thienpondt, and Kris Demuynck. 2020.
\newblock {ECAPA-TDNN:} emphasized channel attention, propagation and
  aggregation in {TDNN} based speaker verification.
\newblock In \emph{Interspeech 2020, 21st Annual Conference of the
  International Speech Communication Association, Virtual Event, Shanghai,
  China, 25-29 October 2020}, pages 3830--3834. {ISCA}.

\bibitem[{Fiscus(1997)}]{fiscus1997post}
Jonathan~G Fiscus. 1997.
\newblock A post-processing system to yield reduced word error rates:
  Recognizer output voting error reduction ({ROVER}).
\newblock In \emph{1997 IEEE Workshop on Automatic Speech Recognition and
  Understanding Proceedings}, pages 347--354. IEEE.

\bibitem[{Gulati et~al.(2020)Gulati, Qin, Chiu, Parmar, Zhang, Yu, Han, Wang,
  Zhang, Wu, and Pang}]{DBLP:conf/interspeech/GulatiQCPZYHWZW20}
Anmol Gulati, James Qin, Chung{-}Cheng Chiu, Niki Parmar, Yu~Zhang, Jiahui Yu,
  Wei Han, Shibo Wang, Zhengdong Zhang, Yonghui Wu, and Ruoming Pang. 2020.
\newblock Conformer: Convolution-augmented transformer for speech recognition.
\newblock In \emph{Interspeech 2020, 21st Annual Conference of the
  International Speech Communication Association, Virtual Event, Shanghai,
  China, 25-29 October 2020}, pages 5036--5040. {ISCA}.

\bibitem[{Hernandez et~al.(2018)Hernandez, Nguyen, Ghannay, Tomashenko, and
  Est{\`{e}}ve}]{DBLP:conf/specom/HernandezNGTE18}
Fran{\c{c}}ois Hernandez, Vincent Nguyen, Sahar Ghannay, Natalia~A. Tomashenko,
  and Yannick Est{\`{e}}ve. 2018.
\newblock {TED-LIUM} 3: Twice as much data and corpus repartition for
  experiments on speaker adaptation.
\newblock In \emph{Speech and Computer - 20th International Conference,
  {SPECOM} 2018, Leipzig, Germany, September 18-22, 2018, Proceedings}, volume
  11096 of \emph{Lecture Notes in Computer Science}, pages 198--208. Springer.

\bibitem[{Jia et~al.(2019)Jia, Weiss, Biadsy, Macherey, Johnson, Chen, and
  Wu}]{DBLP:conf/interspeech/JiaWBMJCW19}
Ye~Jia, Ron~J. Weiss, Fadi Biadsy, Wolfgang Macherey, Melvin Johnson, Zhifeng
  Chen, and Yonghui Wu. 2019.
\newblock Direct speech-to-speech translation with a sequence-to-sequence
  model.
\newblock In \emph{Interspeech 2019, 20th Annual Conference of the
  International Speech Communication Association}, pages 1123--1127. {ISCA}.

\bibitem[{Kim et~al.(2021)Kim, Kong, and Son}]{DBLP:conf/icml/KimKS21}
Jaehyeon Kim, Jungil Kong, and Juhee Son. 2021.
\newblock Conditional variational autoencoder with adversarial learning for
  end-to-end text-to-speech.
\newblock In \emph{Proceedings of the 38th International Conference on Machine
  Learning, {ICML} 2021, 18-24 July 2021, Virtual Event}, volume 139 of
  \emph{Proceedings of Machine Learning Research}, pages 5530--5540. {PMLR}.

\bibitem[{Kim et~al.(2023)Kim, Wu, Peng, Pan, Sridhar, Han, and
  Watanabe}]{kim2023branchformer}
Kwangyoun Kim, Felix Wu, Yifan Peng, Jing Pan, Prashant Sridhar, Kyu~J Han, and
  Shinji Watanabe. 2023.
\newblock {E-Branchformer}: Branchformer with enhanced merging for speech
  recognition.
\newblock In \emph{2022 IEEE Spoken Language Technology Workshop (SLT)}, pages
  84--91. IEEE.

\bibitem[{Kingma and Ba(2015)}]{DBLP:journals/corr/KingmaB14}
Diederik~P. Kingma and Jimmy Ba. 2015.
\newblock \href {http://arxiv.org/abs/1412.6980} {Adam: {A} method for
  stochastic optimization}.
\newblock In \emph{3rd International Conference on Learning Representations,
  {ICLR} 2015, San Diego, CA, USA, May 7-9, 2015, Conference Track
  Proceedings}.

\bibitem[{Kong et~al.(2020)Kong, Kim, and Bae}]{DBLP:conf/nips/KongKB20}
Jungil Kong, Jaehyeon Kim, and Jaekyoung Bae. 2020.
\newblock {HiFi-GAN}: Generative adversarial networks for efficient and high
  fidelity speech synthesis.
\newblock In \emph{Advances in Neural Information Processing Systems 33: Annual
  Conference on Neural Information Processing Systems 2020, NeurIPS 2020,
  December 6-12, 2020, virtual}.

\bibitem[{Lee et~al.(2022)Lee, Chen, Wang, Gu, Popuri, Ma, Polyak, Adi, He,
  Tang, Pino, and Hsu}]{DBLP:conf/acl/LeeCWGPMPAHTPH22}
Ann Lee, Peng{-}Jen Chen, Changhan Wang, Jiatao Gu, Sravya Popuri, Xutai Ma,
  Adam Polyak, Yossi Adi, Qing He, Yun Tang, Juan Pino, and Wei{-}Ning Hsu.
  2022.
\newblock Direct speech-to-speech translation with discrete units.
\newblock In \emph{Proceedings of the 60th Annual Meeting of the Association
  for Computational Linguistics (Volume 1: Long Papers), {ACL} 2022}, pages
  3327--3339. Association for Computational Linguistics.

\bibitem[{Lewis et~al.(2020)Lewis, Liu, Goyal, Ghazvininejad, Mohamed, Levy,
  Stoyanov, and Zettlemoyer}]{DBLP:conf/acl/LewisLGGMLSZ20}
Mike Lewis, Yinhan Liu, Naman Goyal, Marjan Ghazvininejad, Abdelrahman Mohamed,
  Omer Levy, Veselin Stoyanov, and Luke Zettlemoyer. 2020.
\newblock {BART:} denoising sequence-to-sequence pre-training for natural
  language generation, translation, and comprehension.
\newblock In \emph{Proceedings of the 58th Annual Meeting of the Association
  for Computational Linguistics, {ACL} 2020, Online, July 5-10, 2020}, pages
  7871--7880. Association for Computational Linguistics.

\bibitem[{Lim and Oppenheim(1979)}]{lim1979enhancement}
Jae~Soo Lim and Alan~V Oppenheim. 1979.
\newblock Enhancement and bandwidth compression of noisy speech.
\newblock \emph{Proceedings of the IEEE}, 67(12):1586--1604.

\bibitem[{Lison et~al.(2018)Lison, Tiedemann, and
  Kouylekov}]{DBLP:conf/lrec/LisonTK18}
Pierre Lison, J{\"{o}}rg Tiedemann, and Milen Kouylekov. 2018.
\newblock Opensubtitles2018: Statistical rescoring of sentence alignments in
  large, noisy parallel corpora.
\newblock In \emph{Proceedings of the Eleventh International Conference on
  Language Resources and Evaluation, {LREC} 2018, Miyazaki, Japan, May 7-12,
  2018}. European Language Resources Association {(ELRA)}.

\bibitem[{Liu et~al.(2021)Liu, Xu, Wang, Chen, Li, Tan, Li, He, and
  Zhao}]{DBLP:journals/corr/abs-2110-12612}
Yanqing Liu, Zhihang Xu, Gang Wang, Kuan Chen, Bohan Li, Xu~Tan, Jinzhu Li, Lei
  He, and Sheng Zhao. 2021.
\newblock {DelightfulTTS}: The microsoft speech synthesis system for blizzard
  challenge 2021.
\newblock \emph{CoRR}, abs/2110.12612.

\bibitem[{Liu et~al.(2020)Liu, Gu, Goyal, Li, Edunov, Ghazvininejad, Lewis, and
  Zettlemoyer}]{DBLP:journals/tacl/LiuGGLEGLZ20}
Yinhan Liu, Jiatao Gu, Naman Goyal, Xian Li, Sergey Edunov, Marjan
  Ghazvininejad, Mike Lewis, and Luke Zettlemoyer. 2020.
\newblock Multilingual denoising pre-training for neural machine translation.
\newblock \emph{Trans. Assoc. Comput. Linguistics}, 8:726--742.

\bibitem[{Morise(2017)}]{DBLP:conf/interspeech/Morise17}
Masanori Morise. 2017.
\newblock Harvest: {A} high-performance fundamental frequency estimator from
  speech signals.
\newblock In \emph{Interspeech 2017, 18th Annual Conference of the
  International Speech Communication Association}, pages 2321--2325. {ISCA}.

\bibitem[{Nakamura et~al.(2006)Nakamura, Markov, Nakaiwa, Kikui, Kawai,
  Jitsuhiro, Zhang, Yamamoto, Sumita, and
  Yamamoto}]{DBLP:journals/taslp/NakamuraMNKKJZYSY06}
Satoshi Nakamura, Konstantin Markov, Hiromi Nakaiwa, Gen{-}ichiro Kikui,
  Hisashi Kawai, Takatoshi Jitsuhiro, Jinsong Zhang, Hirofumi Yamamoto,
  Eiichiro Sumita, and Seiichi Yamamoto. 2006.
\newblock The {ATR} multilingual speech-to-speech translation system.
\newblock \emph{{IEEE} Trans. Speech Audio Process.}, 14(2):365--376.

\bibitem[{Ojala and Garriga(2010)}]{DBLP:journals/jmlr/OjalaG10}
Markus Ojala and Gemma~C. Garriga. 2010.
\newblock Permutation tests for studying classifier performance.
\newblock \emph{J. Mach. Learn. Res.}, 11:1833--1863.

\bibitem[{Panayotov et~al.(2015)Panayotov, Chen, Povey, and
  Khudanpur}]{DBLP:conf/icassp/PanayotovCPK15}
Vassil Panayotov, Guoguo Chen, Daniel Povey, and Sanjeev Khudanpur. 2015.
\newblock Librispeech: An {ASR} corpus based on public domain audio books.
\newblock In \emph{2015 {IEEE} International Conference on Acoustics, Speech
  and Signal Processing, {ICASSP} 2015, South Brisbane, Queensland, Australia,
  April 19-24, 2015}, pages 5206--5210. {IEEE}.

\bibitem[{Park et~al.(2019)Park, Chan, Zhang, Chiu, Zoph, Cubuk, and
  Le}]{DBLP:conf/interspeech/ParkCZCZCL19}
Daniel~S. Park, William Chan, Yu~Zhang, Chung{-}Cheng Chiu, Barret Zoph,
  Ekin~D. Cubuk, and Quoc~V. Le. 2019.
\newblock {SpecAugment}: {A} simple data augmentation method for automatic
  speech recognition.
\newblock In \emph{Interspeech 2019, 20th Annual Conference of the
  International Speech Communication Association, Graz, Austria, 15-19
  September 2019}, pages 2613--2617. {ISCA}.

\bibitem[{Radford et~al.(2018)Radford, Narasimhan, Salimans, Sutskever
  et~al.}]{radford2018improving}
Alec Radford, Karthik Narasimhan, Tim Salimans, Ilya Sutskever, et~al. 2018.
\newblock Improving language understanding by generative pre-training.

\bibitem[{Rousseau et~al.(2012)Rousseau, Del{\'{e}}glise, and
  Est{\`{e}}ve}]{DBLP:conf/lrec/RousseauDE12}
Anthony Rousseau, Paul Del{\'{e}}glise, and Yannick Est{\`{e}}ve. 2012.
\newblock {TED-LIUM:} an automatic speech recognition dedicated corpus.
\newblock In \emph{Proceedings of the Eighth International Conference on
  Language Resources and Evaluation, {LREC} 2012, Istanbul, Turkey, May 23-25,
  2012}, pages 125--129. European Language Resources Association {(ELRA)}.

\bibitem[{Shi et~al.(2021)Shi, Bu, Xu, Zhang, and
  Li}]{DBLP:conf/interspeech/ShiBXZL21}
Yao Shi, Hui Bu, Xin Xu, Shaoji Zhang, and Ming Li. 2021.
\newblock {AISHELL-3:} {A} multi-speaker mandarin {TTS} corpus.
\newblock In \emph{Interspeech 2021, 22nd Annual Conference of the
  International Speech Communication Association, Brno, Czechia, 30 August - 3
  September 2021}, pages 2756--2760. {ISCA}.

\bibitem[{Sohn et~al.(1999)Sohn, Kim, and Sung}]{DBLP:journals/spl/SohnKS99}
Jongseo Sohn, Nam~Soo Kim, and Wonyong Sung. 1999.
\newblock A statistical model-based voice activity detection.
\newblock \emph{{IEEE} Signal Process. Lett.}, 6(1):1--3.

\bibitem[{Song et~al.(2022{\natexlab{a}})Song, Xue, Wang, Cong, Zhang, Xie,
  Yang, Zhang, and Su}]{DBLP:conf/iscslp/SongXWCZXYZS22}
Kun Song, Heyang Xue, Xinsheng Wang, Jian Cong, Yongmao Zhang, Lei Xie, Bing
  Yang, Xiong Zhang, and Dan Su. 2022{\natexlab{a}}.
\newblock {AdaVITS}: Tiny {VITS} for low computing resource speaker adaptation.
\newblock In \emph{13th International Symposium on Chinese Spoken Language
  Processing, {ISCSLP} 2022, Singapore, December 11-14, 2022}, pages 319--323.
  {IEEE}.

\bibitem[{Song et~al.(2022{\natexlab{b}})Song, Zhang, Lei, Cong, Li, Xie, He,
  and Bai}]{DBLP:journals/corr/abs-2211-01087}
Kun Song, Yongmao Zhang, Yi~Lei, Jian Cong, Hanzhao Li, Lei Xie, Gang He, and
  Jinfeng Bai. 2022{\natexlab{b}}.
\newblock {DSPGAN:} a {GAN}-based universal vocoder for high-fidelity {TTS} by
  time-frequency domain supervision from {DSP}.
\newblock \emph{CoRR}, abs/2211.01087.

\bibitem[{Ye et~al.(2022)Ye, Zhao, Ko, Meng, Wang, Wang, and
  Cao}]{DBLP:journals/corr/abs-2204-03939}
Rong Ye, Chengqi Zhao, Tom Ko, Chutong Meng, Tao Wang, Mingxuan Wang, and Jun
  Cao. 2022.
\newblock {GigaST}: {A} 10, 000-hour pseudo speech translation corpus.
\newblock \emph{CoRR}, abs/2204.03939.

\bibitem[{Zhang et~al.(2022)Zhang, Xue, Li, Xie, Guo, Zhang, and
  Gong}]{DBLP:journals/corr/abs-2211-02903}
Yongmao Zhang, Heyang Xue, Hanzhao Li, Lei Xie, Tingwei Guo, Ruixiong Zhang,
  and Caixia Gong. 2022.
\newblock Visinger 2: High-fidelity end-to-end singing voice synthesis enhanced
  by digital signal processing synthesizer.
\newblock \emph{CoRR}, abs/2211.02903.

\end{thebibliography}

\appendix
\section{Appendix}
We present the official results, which include our submitted system and those of other teams.   As shown in Table~\ref{human}, our system ranks 1st in speech quality score and 2nd in translation quality score.   By equally weighting translation quality and speech quality, our submitted system achieves the highest overall score in human evaluation.   Although the organizers provide both automatic and human evaluation scores, the systems are ranked based on human evaluation.   Consequently, our submitted system \textbf{ranks 1st} in the S2ST task of the IWSLT 2023 evaluation campaign.   Additionally, as illustrated in Table~\ref{auto}, we rank 2nd and closely follow the 1st place in automatic evaluation, which evaluates translation accuracy.   Our system employs zero-shot voice cloning, which may result in a slight loss of sound quality and speech clarity.   We believe our automatic evaluation results could be better without using zero-shot voice cloning.   However, this trade-off allows us to achieve a significant improvement in speaker timbre similarity and naturalness.

\begin{table*}[tp]
\centering
    \vspace{-5pt}
	\caption{Official results of the human evaluation.}
 \vspace{-6pt}
\resizebox{0.95\linewidth}{!}{ 
\begin{tabular}{lccc}
    \toprule
 \textbf{System}  & \textbf{Translation Quality Score}  & \textbf{Speech Quality Score} & \textbf{Overall} \\ \midrule
\midrule
\textit{Cascade Systems} & & \\
\midrule
\textbf{Our Submitted System} & 3.70 & 3.98 & 3.84 \\
XIAOMI & 3.72 & 3.67 & 3.70 \\
 HW-TSC & 3.58 & 3.75 & 3.67 \\
 MINETRANS\_Cascade & 3.16 & 3.26 & 3.21 \\
 KU & 2.92 & 3.01 & 2.97 \\

\midrule
\midrule
\textit{E2E Systems} & & \\
\midrule
 MINETRANS\_E2E(contrastive2) & 3.58 & 3.50 & 3.54 \\

\bottomrule
\end{tabular}
}

\label{human}
\renewcommand{\thefigure}{1}
\end{table*}

\vspace{50pt}

\begin{table*}[]
	\caption{Official results of the automatic evaluation.}
 \vspace{-6pt}
\resizebox{\linewidth}{!}{
\begin{tabular}{lcccccccccccc}
\hline
\multicolumn{1}{l|}{\textbf{System}}               & \multicolumn{4}{c|}{\textbf{Test-primary}}                                                                            & \multicolumn{4}{c|}{\textbf{Test-expanded}}                                                                           & \multicolumn{4}{c}{\textbf{Overall}}                                                                                 \\
\multicolumn{1}{l|}{Ref}                           & \multicolumn{1}{l}{{BLEU}} & \multicolumn{1}{l}{chrF} & \multicolumn{1}{l}{COMET} & \multicolumn{1}{l|}{SEScore2} & \multicolumn{1}{l}{{BLEU}} & \multicolumn{1}{l}{chrF} & \multicolumn{1}{l}{COMET} & \multicolumn{1}{l|}{SEScore2} & \multicolumn{1}{l}{{BLEU}} & \multicolumn{1}{l}{chrF} & \multicolumn{1}{l}{COMET} & \multicolumn{1}{l}{SEScore2} \\ \hline
\textit{Cascade Systems}                                     & \multicolumn{1}{l}{}           & \multicolumn{1}{l}{}     & \multicolumn{1}{l}{}      & \multicolumn{1}{l}{}          & \multicolumn{1}{l}{}           & \multicolumn{1}{l}{}     & \multicolumn{1}{l}{}      & \multicolumn{1}{l}{}          & \multicolumn{1}{l}{}           & \multicolumn{1}{l}{}     & \multicolumn{1}{l}{}      & \multicolumn{1}{l}{}         \\ \hline
\multicolumn{1}{l|}{XIAOMI}                        & 47.9                           & 41.0                     & 79.91                     & \multicolumn{1}{c|}{-12.27}   & 34.5                           & 29.2                     & 79.07                     & \multicolumn{1}{c|}{-20.15}   & 38.4                           & 32.3                     & 79.35                     & -17.48                       \\
\multicolumn{1}{l|}{Our Sumbitted System}          & 47.4                           & 40.7                     & 79.90                     & \multicolumn{1}{c|}{-12.21}   & 34.0                           & 28.5                     & 78.68                     & \multicolumn{1}{c|}{-20.23}   & 37.7                           & 31.8                     & 79.09                     & -17.52                       \\
\multicolumn{1}{l|}{HW-TSC}                        & 43.2                           & 36.9                     & 76.96                     & \multicolumn{1}{c|}{-14.23}   & 32.4                           & 27.7                     & 76.43                     & \multicolumn{1}{c|}{-21.61}   & 35.3                           & 30.1                     & 76.61                     & -19.12                       \\
\multicolumn{1}{l|}{KU}                            & 36.7                           & 31.3                     & 69.09                     & \multicolumn{1}{c|}{-17.07}   & 25.0                           & 21.7                     & 67.94                     & \multicolumn{1}{c|}{-25.68}   & 28.2                           & 24.3                     & 68.33                     & -22.77                       \\
\multicolumn{1}{l|}{MINETRANS\_Cascade}            & 33.9                           & 28.6                     & 67.49                     & \multicolumn{1}{c|}{-17.68}   & 24.7                           & 21.5                     & 64.71                     & \multicolumn{1}{c|}{-26.34}   & 27.2                           & 23.4                     & 65.65                     & -23.41                       \\ \hline
\textit{E2E Systems}                                         &                                &                          &                           &                               &                                &                          &                           &                               &                                &                          &                           &                              \\ \hline
\multicolumn{1}{l|}{MINETRANS\_E2E (contrastive2)} & 45.0                           & 38.3                     & 74.83                     & \multicolumn{1}{c|}{-13.62}   & 31.1                           & 26.4                     & 73.28                     & \multicolumn{1}{c|}{-22.03}   & 34.9                           & 29.6                     & 73.81                     & -19.18                       \\
\multicolumn{1}{l|}{MINETRANS\_E2E (contrastive1)} & 44.5                           & 38.0                     & 74.14                     & \multicolumn{1}{c|}{-13.92}   & 31.0                           & 26.4                     & 72.90                     & \multicolumn{1}{c|}{-22.20}   & 34.8                           & 29.5                     & 73.32                     & -19.40                       \\
\multicolumn{1}{l|}{MINETRANS\_E2E (primary)}      & 44.4                           & 38.0                     & 74.40                     & \multicolumn{1}{c|}{-13.86}   & 31.1                           & 26.4                     & 73.00                     & \multicolumn{1}{c|}{-22.12}   & 34.7                           & 29.5                     & 73.47                     & -19.32                       \\ \hline
\end{tabular}
}
\label{auto}
\end{table*}

\end{document}